\documentclass{imsphy}

\usepackage{graphicx}
\usepackage{amsmath}
\usepackage{xcolor}
\usepackage{verbatim}
\usepackage{setspace}
\usepackage{amssymb}
\usepackage{empheq}
\usepackage[square,numbers,sort&compress]{natbib}

\usepackage{hyperref} 

\usepackage{soul}                       

\hypersetup{
    colorlinks=true, 
    linktoc=all,     
    linkcolor=blue,  
    filecolor=blue,
    citecolor=blue,
    urlcolor=blue,
    pdfpagemode=UseOutlines
}

\setstcolor{red}    
\setulcolor{red}    

\begin{document}

\label{chap:article}

\begin{center}
{\Large {\bf Symmetry of the Klein-Gordon Equation under Disformal Transformations with Higher-Order Derivatives}}\\
\vspace{0.3cm}
Yurev Ivan Ross B. Carag and Allan L. Alinea\\
\vspace{0.2cm}
{\it Astrophysics, Particle Physics, and Nuclear Physics Research Cluster\\Institute of  Physics, University of the 
Philippines Los Ba\~nos\\ 4031 College, Los Ba\~nos, Laguna, Philippines}
\vspace{0.5cm}

\begin{minipage}{5in}
{\bf Abstract}\\
Scalar-Tensor theories extend General Relativity (GR) by including both a scalar field and a metric tensor in describing gravity. Although the matter sector is minimally coupled to the metric in GR, the scalar field coupling to matter may be adjusted by means of Disformal Transformations. One of the most general of these transformations, formulated by Takahashi, Motohashi, and Minamitsuji, include arbitrarily high-order derivatives of the scalar field and are designed to map between two ghost-free Higher-order Scalar-Tensor theories. We used these transformations to determine the most general conditions for the disformal form-invariance of the Klein-Gordon equation for the scalar field. In the second-order case, we identified a family of disformally-related metrics that result in the same propagation solutions. Additionally, higher-order cases could comply with theoretical and phenomenological constraints alongside KG-invariance. Finally, we investigated the notion of disformal transformations as ``deformations'' of the spacetime manifold. \\

\end{minipage}
\end{center}

\vspace{0.5cm}

\noindent{\large {\bf 1 Introduction}}

\noindent Metric theories of gravity, like GR, identify gravity as the curvature of a spacetime manifold with metric $g_{\mu \nu}$ whose geodesics are the free-fall paths of test objects. Scalar-Tensor theories, on the other hand, associate gravitation with not only with a background metric $g_{\mu\nu}$, but also a scalar field $\phi$. Because of the added scalar field in the description of gravity, matter does not necessarily follow the geodesics of the background spacetime  \cite{bekenstein1993}. This means that one could interpret these theories in two different ways, namely the Einstein and Jordan frames. The Jordan (physical) frame adds the scalar field to the description of the geometry of spacetime, while the Einstein (background) frame identifies the scalar field as another source of curvature. Mainstream graduate texts on Relativity usually suggest that these two frames are connected via a conformal transformation \cite{gron2007}\cite{carroll2003}. However, for other Scalar-Tensor theories, like Horndeski or Degenerate Higher-Order Scalar-Tensor (DHOST) theories, the connection between the background and physical frames is through a disformal transformation. As of this writing, the most general disformal transformations are those formulated by Takahashi, Motohashi, and Minamitsuji which are meant to map one DHOST theory to another without introducing Ostrogradsky ghosts \cite{takahashi2022}. In this paper, we shall call these the ``TMM Disformal Transformations.''

In this study, we perform the TMM Disformal Transformation to the Klein-Gordon equation, determine the necessary and sufficient conditions to satisfy disformal KG-invariance, and examine the effect of the constrained TMM disformal transformation on the energy-momentum tensor associated with the Klein-Gordon action.
\vspace{0.5cm}

\noindent{\large{\bf 2 The disformal transformation}}

\noindent The Takahashi-Motohashi-Minamitsuji (TMM) class of disformal transformations may be enacted by using a transformed metric given by
\begin{equation}\label{apeq:g_hat}
\widehat{g}_{\mu\nu}=Ag_{\mu\nu}+C\phi_{;\mu}\phi_{;\nu}+2D\phi_{(;\mu}X_{;\nu)}+EX_{;\mu}X_{;\nu}
\end{equation}
where $A=A(\phi,X,Y,Z)$ is the conformal factor and $C$, $D$, and $E$, called disformal factors, are all functionals of $\phi,X,Y,Z$. The parentheses enclosing two indices denotes symmetrization, namely $\phi_{(;\mu}X_{;\nu)}=\frac{1}{2}(\phi_{;\mu}X_{;\nu}+\phi_{;\nu}X_{;\mu})$ The metric shown in (\ref{apeq:g_hat}) which is up to the second-order derivative of $\phi$ shall be referred to as the \emph{default} TMM disformal metric. Cases with higher-order derivatives of $\phi$ beyond $X_{;\mu}$ shall be considered extensions of the default case.

The expression for the disformal metric in (\ref{apeq:g_hat}) is written in terms of components. We may also express it in the following way:
\begin{align}\label{apeq:g_hat_mat_comp}
\widehat{g}_{\mu\nu}&=Ag_{\mu\nu}+
 \begin{pmatrix}
 \phi_{;\mu} & X_{;\mu} \\
\end{pmatrix} \begin{pmatrix}
 C & D \\
 D & E \\
\end{pmatrix} \begin{pmatrix}
 \phi_{;\nu}\\X_{;\nu}
\end{pmatrix} 
\\
\label{apeq:g_hat_mat}
\widehat{\boldsymbol{g}}&=A{\boldsymbol g}+\boldsymbol K^T\boldsymbol B\boldsymbol K
\\
\label{apeq:g_hat_mat_2}
\widehat{\boldsymbol{g}}&=A{\boldsymbol g}+\boldsymbol \xi^T\boldsymbol H\boldsymbol \xi
\\
\label{apeq:g_hat_diag_4}
\widehat{\boldsymbol{g}}&=A{\boldsymbol g}+\sum_{i=1}^{k-r}\lambda_i\xi^{(i)}_{\mu}\xi^{(i)}_{\nu}
\end{align}
for an $n$-dimensional spacetime, with $k$ of the field derivatives $(\phi_{;\mu},X_{;\mu},Y_{;\mu}, \text{etc.})$. The matrix $\boldsymbol B$ is the coefficient matrix containing $C,D,E,...$ and $\boldsymbol H$ is its diagonalized matrix containing the $k-r$ nonzero eigenvalues $\lambda_i$ of $\boldsymbol B$. To preserve the utility of the metric as a definition of a commutative scalar product, $\boldsymbol B$ must be symmetric. 
\vspace{0.5cm}

\noindent{\large{\bf 3 Transforming the d'Alembertian}}

\noindent Transforming the Klein-Gordon equation involves switching from the metric  $g_{\mu \nu}$ to the hatted metric $\widehat{g}_{\mu \nu}$. This switch means that the mass $m$ might also need to be redefined. Hence, the transformation is mathematically carried out as
\begin{align}
(\square - m^2)\phi\rightarrow (\widehat{\square} - \widehat{m}^2)\phi,
\label{apeq:KG_transf}
\end{align}
where $\widehat{m}$ is the mass in the hatted frame, ${\square}\phi$ is given by
\begin{align}
{\square}\phi &\equiv\frac{({g}^{\mu \nu}\sqrt{-{g}} \phi_{,\mu})_{,\nu}}{\sqrt{-{g}}},
\label{apeq:dAlemDef1}
\end{align}
and $\widehat{\square}\phi$ is the same form as \eqref{apeq:dAlemDef1} but with $g_{\mu\nu}$ replaced with $\widehat{g}_{\mu\nu}$. To transform the Klein-Grodon equation, we need to calculate the transformation rule of the $\square{\phi}$ , which requires finding expressions for raised metric $\widehat{g}^{\mu \nu}$ and determinant $\widehat{g}$ in terms of their unhatted counterparts. Using the Woodbury Matrix formula \cite{higham2002} for $\widehat{g}^{\mu\nu}$  and Matrix Determinant lemma \cite{brookes2020} for $g$, the transformation of the d'Alembertian was found to be
 \begin{align}
\widehat{\square}\phi=&\frac{\square \phi}{A}+\frac{1}{2A}g^{\mu\nu}\phi_{,\mu}\bigg[\ln{\bigg(A^{n-2}\frac{\lambda_1 \lambda_2 \cdots\lambda_{k-r}}{\zeta^1 \zeta^2  \cdots\zeta^{k-r}}\bigg)}\bigg]_{,\nu}
\nonumber
\\
&-\frac{1}{A^2}\sum_{i=1}^{k-r}\zeta^{i}\sigma_{(i)}\bigg{\{}\square{Q_{(i)}}+\frac{1}{2}g^{\mu\nu}Q_{(i),\mu}\bigg[\ln{\bigg(A^{n-4}\frac{\lambda_1 \lambda_2 \cdots\lambda_{k-r}}{\zeta^1 \zeta^2  \cdots\zeta^{i-1}\zeta^{i+1}\cdots\zeta^{k-r}}\bigg)}\bigg]_{,\nu}
\bigg{\}},
\label{apeq:hatted_dAlem_unconstrained}
\end{align}
where the $\lambda_i$ and $\zeta^i$ are the nonzero eigenvalues of $\boldsymbol B$ and its corresponding inverse coefficient matrix, respectively; $Q_{(i)}$ is a functional of $\phi,X,Y,Z,...$ and $\sigma_{(i)}=\chi_{(i)}^{\mu}\phi_{,\mu}$, with $\chi_{(i)\mu}$ being the matrix inverse counterpart of $\xi^{(i)}_{\mu}$ from \eqref{apeq:g_hat_mat_2}.
\vspace{0.5cm}

\noindent{\large{\bf 4 The KG-Constrained TMM Disformal Transformation and the Energy-Momentum Tensor}}

\noindent In order for the Klein-Gordon equation to be form-invariant under disformal transformations and remove ghost terms, i.e. d'Alemberatians of higher-order derivatives, the last two terms on the right-hand side of \eqref{apeq:hatted_dAlem_unconstrained} should each vanish, leading to two conditions for KG form-invariance:
\begin{align}
\label{apeq:conditions}
g^{\mu\nu}\xi^{(i)}_{\mu}\phi_{,\mu}=0
\quad\mathrm{and}\quad\widehat{g}=A^2g.
\end{align}
Implementing these two conditions for the default TMM case forces the disformal factors to be related to $A$ in the following manner:
\begin{align} 
     \label{apeq:C}
     C&=\frac{Y^2(A^{3-n}-A)}{2X(Y^2+2XZ)},\\
     \label{apeq:D}
     D&=\frac{Y(A^{3-n}-A)}{(Y^2+2XZ)},\\
     \label{apeq:E}
     E&=\frac{2X(A^{3-n}-A)}{(Y^2+2XZ)}.  
\end{align}
Using relations \eqref{apeq:C} through \eqref{apeq:E}, we can create a continuous set $\mathcal{G}$ of metrics, parametrized by $A$, associated with manifolds wherein the propagation of $\phi$ dictated by the Klein-Gordon equation would be the same in the massless case. If the scalar field is massive, then the mass would transform as $\widehat m=\frac{m}{\sqrt{A}}$. The conditions in \eqref{apeq:conditions} provide a total of $\frac{r(r-1)}{2}+k+1$ total constraints for $\frac{k(k+1)}{2}$ disformal factors. Therefore, the $k=2$ case is exactly constrained by KG-invariance. Meanwhile, the higher-order $(k>2)$ cases are underconstrained. Hence, this class of disformal transformations offers the possibility of being tuned to meet other theoretical or phenomenological requirements alongside KG-invariance. 

The mass transformation presages a mass-energy exchange between the scalar field and the warping of the manifold through disformal transformation. We further explore this idea through the energy-momentum tensor of $\phi$, which is given by $T_{\mu\nu}=\phi_{,\mu}\phi_{,\nu}-g_{\mu\nu}(X-\frac{1}{2}m^2\phi^2)$ obtained from the Einstein-Hilbert definition. Performing a disformal transformation constrained by \eqref{apeq:conditions} gives an energy-momentum tensor in the hatted frame of the form $\widehat{T}_{\mu\nu}=T_{\mu\nu}+N_{\mu\nu}$, where $N_{\mu\nu}$ is proportional to the disformal terms, as well as $m$ and $X$, hence it can be interpreted as an induced curvature source due to disformal transformation. Because of KG-invariance, both tensors are conserved currents in their own frames, implying $\widehat{\nabla}^{\mu}T_{\mu\nu}=-\widehat{\nabla}^{\mu}N_{\mu\nu}$, which suggests an exchange relation between energy-momentum and deformations of the spacetime manifold through the disformal transformation.
\vspace{0.5cm}

\noindent{\large{\bf 5 Summary and conclusion}}

\noindent In this paper, we were able to determine how the form of the Klein-Gordon equation changes under a TMM Disformal Transformation with arbitrarily-high order derivatives using the framework of linear algebra. We then identified \eqref{apeq:conditions} as the set of general conditions, along with the mass transformation $\widehat{m}=\frac{m}{\sqrt{A}}$ to maintain KG-invariance, with the $k=2$ case sufficiently constrained but the $k>2$ cases under-constrained. Lastly, we were able to explore the notion of exchange between energy-momentum of matter and the warping of spacetime through the behavior of the energy-momentum tensor under the KG-constrained disformal transformations.

\bibliographystyle{plain}

\end{document}